\documentstyle{article}

\newcommand{\be}{\begin{equation}}
\newcommand{\ee}{\end{equation}}
\newcommand{\ba}{\begin{eqnarray}}
\newcommand{\ea}{\end{eqnarray}}
\newcommand{\no}{\nonumber\\}

\newcommand{\imag}{\mbox{Im}}
\newcommand{\real}{\mbox{Re}}

\title{Comment on ``Indirect CP violation in the B-system''}

\author{L.\ Lavoura \\
{\small Universidade T\'ecnica de Lisboa} \\
{\small Centro de F\'\i sica das Interac\c c\~oes Fundamentais} \\
{\small Instituto Superior T\'ecnico, P-1096 Lisboa Codex, Portugal} \\
{\small Email i009@beta.ist.utl.pt. Telephone (+351 1) 841 9093.
Fax (+351 1) 841 9143.}}

\date{24 June 1998}

\begin{document}
\maketitle

\begin{abstract}
In a recent paper,
Ba\~nuls and Bernab\'eu have claimed the existence
of a new form of indirect CP violation
which would have its most prominent manifestation
in the $B_d$-$\bar B_d$ system.
I analyse this claim in detail.
I emphasize the fact that it is necessary to take into account
the precise experimental method if one is to identify correctly
the weak phase that one is about to measure.
\end{abstract}

\vspace{5mm}

\section{Introduction}

In a recent paper \cite{banuls},
Ba\~nuls and Bernab\'eu (BB) have claimed that:
\begin{enumerate}
\item It is possible to define a complex rephasing-invariant
parameter $\epsilon$ in the mixing of each neutral-meson system
(for instance,
the $K$-$\bar K$ or the $B_d$-$\bar B_d$ system),
such that $\real\, \epsilon$ and $\imag\, \epsilon$ are two independent,
measurable CP-violating quantities.
\item While
\be
\frac{2 \real\, \epsilon}{1 + \left| \epsilon \right|^2} = \frac
{\left| p \right|^2 - \left| q \right|^2}
{\left| p \right|^2 + \left| q \right|^2}
\label{real de epsilon}
\ee
is the usual parameter of indirect CP violation,
which is experimentally known
to be no larger than $10^{-1}$ in the $B_d$-$\bar B_d$ system \cite{cleo},
$\imag\, \epsilon$ corresponds to a new form of indirect CP violation,
which should be particularly large in that system.
BB have predicted that
\be
-0.74 < \frac{2 \imag\, \epsilon}{1 + \left| \epsilon \right|^2} < -0.36
\label{BB's value}
\ee
in the $B_d$-$\bar B_d$ system.
\item By observing the time dependence of the flavour-specific decays of states
which at initial time were tagged to be one of the CP eigenstates
of the neutral-meson system,
one may construct an asymmetry which allows one to determine
both the real and the imaginary parts of $\epsilon$.
\end{enumerate}

The results advertised by BB are surprising for many reasons.
BB not only claim the existence of an hitherto unnoticed,
completely new form of CP violation in the mixing of neutral mesons,
they also suggest an experiment to measure it,
and they are even able to predict its standard-model value.
The prediction is baffling
in that it is based on nothing more than the well-known
standard-model box diagram for $M_{12}$ in the $B_d$-$\bar B_d$ system.

These amazing results have entailed me to analyse BB's work in detail.
I found that BB's interpretation of some calculations is mostly wrong.

In particular,
the prediction in eqn~(\ref{BB's value})
basically follows from the identification
\be
\frac{2 \imag\, \epsilon}{1 + \left| \epsilon \right|^2}
= - \frac{\imag \left[ \left( V_{tb} V_{td}^\ast \right)^2 \right]}
{\left| V_{tb} V_{td} \right|^2},
\label{the root of the evil}
\ee
where $V$ is the Cabibbo--Kobayashi--Maskawa mixing matrix.
Equation~(\ref{the root of the evil}) arouses suspicion,
because the combination $\left( V_{tb} V_{td}^\ast \right)^2$
is not invariant under a rephasing of the $d$- and $b$-quark fields,
and therefore the imaginary part of that combination is not measurable.
Indeed,
I find that eqn~(\ref{the root of the evil})
neglects the CP-transformation phases of the quarks,
a procedure which,
although commonplace in the literature,
is illegitimate.

Moreover,
BB have explicitly claimed that the product
\be
\langle f|T|B_+ \rangle^\ast \langle f|T|B_- \rangle
\ee
---where $B_+$ and $B_-$ are the CP eigenstates
of the $B_d$-$\bar B_d$ system---is not phase-convention-dependent;
however,
it is clear that independent rephasings of $|B_+\rangle$
and of $|B_-\rangle$ do change the phase of that product.
This fact suggests that,
when BB write that $\epsilon$ is rephasing-invariant,
they are not taking into account the freedom
that one has to {\em independently} rephase {\em all} the kets.
This suspicion proves to be true.

\section{The CP-violating asymmetry}

It is convenient to start by analysing BB's proposal
for the measurement of a well-defined time-dependent CP-violating asymmetry.
The proposed experiment
basically consists in the following.\footnote{For definiteness
I always work in terms of the $B_d$-$\bar B_d$ system,
which is the one for which BB's ``discoveries''
are supposed to be most important.}
One uses a state which,
at initial time $t=0$,
is a coherent superposition
of $|B_d\rangle$ and $|\bar B_d\rangle$
in which there is an equal probability
of finding $B_d$ and of finding $\bar B_d$.
One may write such a state,
in all generality,
as
\be
|B_+\rangle = \frac{e^{i \alpha_+}}{\sqrt{2}}
\left( |B_d\rangle + e^{i \zeta} |\bar B_d\rangle \right).
\label{b+}
\ee
This state evolves into $|B_+(t)\rangle$ at proper time $t > 0$.
One measures the probabilities of finding $B_d$ and of finding $\bar B_d$
in $B_+(t)$ and one constructs the corresponding asymmetry:
\be
A_+^{\rm CP}(t) \equiv \frac
{\left| \langle B_d|B_+(t) \rangle \right|^2 -
\left| \langle \bar B_d|B_+(t) \rangle \right|^2}
{\left| \langle B_d|B_+(t) \rangle \right|^2 +
\left| \langle \bar B_d|B_+(t) \rangle \right|^2}.
\label{definition of A+CP}
\ee
Obviously,
$A_+^{\rm CP}(t) \neq 0$ represents a violation of CP:
if an initial state in which there is an equal probability
of finding a particle and its antiparticle
evolves into a final state in which the probabilities
of finding the particle and the antiparticle are different,
then CP is violated.\footnote{The probabilities
in eqn~(\ref{definition of A+CP}) may be measured
by observing the decays into flavour-specific modes,
like for instance the semileptonic modes $l^\pm X^\mp$,
of $B_+(t)$.
When doing this one must assume that
$\left| \langle l^+ X^-|T|B_d \rangle \right| =
\left| \langle l^- X^+|T|\bar B_d \rangle \right| $;
this equality follows from CPT invariance.}
The asymmetry $A_+^{\rm CP}(t)$
(or its time-integrated version,
which of course contains less information)
is the observable whose measurement has been proposed by BB.

In order to compute $A_+^{\rm CP}(t)$
it is necessary to use the eigenstates of mass,
which I write as
\be
\begin{array}{rcl}
|B_1\rangle \!\!\!&=&\!\!\! p_1 |B_d\rangle + q_1 |\bar B_d\rangle,
\\*[1mm]
|B_2\rangle \!\!\!&=&\!\!\! p_2 |B_d\rangle - q_2 |\bar B_d\rangle.
\end{array}
\label{the eigenstates}
\ee
I assume these states to be normalized:
$\left| p_1 \right|^2 + \left| q_1 \right|^2
= \left| p_2 \right|^2 + \left| q_2 \right|^2 = 1$.
It follows from CPT invariance in the mixing that
\be
\frac{q_1}{p_1} = \frac{q_2}{p_2} \equiv \frac{q}{p}
= \sqrt{\frac{1 - \delta}{1 + \delta}} e^{i \chi},
\ee
where
\be
\delta = \frac{2 \real\, \epsilon}{1 + \left| \epsilon \right|^2}
\ee
---see eqn~(\ref{real de epsilon})---is the usual T- and CP-violating quantity.
Usually one makes a convention for the relative phase
of $|B_1\rangle$ and $|B_2\rangle$
such that $p_1 = p_2 \equiv p$ and $q_1 = q_2 \equiv q$.

The phase
\be
\theta \equiv \zeta - \arg{\frac{q}{p}}
\label{theta}
\ee
is invariant under a change of the relative phase
of $|B_d\rangle$ and $|\bar B_d\rangle$,
contrary to what happens with the phases $\zeta$ and $\arg{q/p}$
separately---{\it cf.} eqns~(\ref{b+}) and (\ref{the eigenstates}).

The states $|B_k\rangle$ ($k = 1, 2$) have exponential evolution laws:
$|B_k(t)\rangle = \exp \left( - i \lambda_k t \right) |B_k\rangle$
with $\lambda_k = m_k - (i/2) \gamma_k$.
Defining $\Delta m \equiv m_2 - m_1$
and $\Delta \Gamma \equiv \gamma_2 - \gamma_1$,
one finds
\ba
A_+^{\rm CP}(t) \!\!\!&=&\!\!\!
\left[
\delta \cosh{\left( \Delta \Gamma t/2 \right) }
+ \delta \sqrt{1 - \delta^2} \cos{\theta}
\sinh{\left( \Delta \Gamma t/2 \right) }
\right.
\no         \!\!\!& &\!\!\!
\left.
- \delta \cos{\left( \Delta m t \right) }
- \sqrt{1 - \delta^2} \sin{\theta} \sin{\left( \Delta m t \right) }
\right]
\no         \!\!\!& &\!\!\!
\times \left[
\cosh{\left( \Delta \Gamma t/2 \right) }
+ \sqrt{1 - \delta^2} \cos{\theta} \sinh{\left( \Delta \Gamma t/2 \right) }
\right.
\no         \!\!\!& &\!\!\!
\left.
- \delta^2 \cos{\left( \Delta m t \right) }
- \delta \sqrt{1 - \delta^2} \sin{\theta} \sin{\left( \Delta m t \right) }
\right]^{-1}.
\label{a+cp(t)}
\ea
By measuring $A_+^{\rm CP}(t)$ one may in principle find $\delta$ and $\theta$.
It is clear from eqn~(\ref{a+cp(t)}) that
$\sin \theta \neq 0$ represents CP violation,
just as $\delta \neq 0$.
Now,
we know that $\delta$ is small,
and therefore we may write
\be
A_+^{\rm CP}(t) \approx
\frac{\delta \left[ \cosh{\left( \Delta \Gamma t/2 \right) }
+ \cos{\theta} \sinh{\left( \Delta \Gamma t/2 \right) }
- \cos{\left( \Delta m t \right) } \right]
- \sin{\theta} \sin{\left( \Delta m t \right) }}
{\cosh{\left( \Delta \Gamma t/2 \right) }
+ \cos{\theta} \sinh{\left( \Delta \Gamma t/2 \right) }
- \delta \sin{\theta} \sin{\left( \Delta m t \right) }}.
\ee
Notice however that it is illegitimate to assume $\sin{\theta}$ to be small;
in general,
$\theta$ depends on $\zeta$,
{\it i.e.},
it depends on the initial state $|B_+\rangle$ that one uses.
Therefore,
$\theta$ may be large or small depending on the particular $B_+$.
It is unwarranted to assume $\theta$
to be of the same order of magnitude as $\delta$,
like BB have done.

\section{Definition of $\epsilon$}

One may unify
the two real CP-violating quantities $\delta$ and $\sin{\theta}$
in a single complex CP-violating parameter $\epsilon$
by means of a simple---albeit meaningless---trick.
Let us consider another coherent superposition
of $|B_d\rangle$ and $|\bar B_d\rangle$
in which the probabilities of finding $B_d$
and of finding $\bar B_d$ are equal
and which is orthogonal to $|B_+\rangle$:
\be
|B_-\rangle = \frac{e^{i \alpha_-}}{\sqrt{2}}
\left( |B_d\rangle - e^{i \zeta} |\bar B_d\rangle \right).
\label{b-}
\ee
We use,
together with BB,
$p_1 = p_2 \equiv p$ and $q_1 = q_2 \equiv q$.
Now,
if one writes
\be
\begin{array}{rcl}
|B_1\rangle \!\!\!&=&\!\!\!
{\displaystyle \frac{1}{\sqrt{1 + \left| \epsilon \right|^2}}
\left( |B_+\rangle + \epsilon |B_-\rangle \right),}
\\*[3mm]
|B_2\rangle \!\!\!&=&\!\!\!
{\displaystyle \frac{1}{\sqrt{1 + \left| \epsilon \right|^2}}
\left( |B_-\rangle + \epsilon |B_+\rangle \right),}
\end{array}
\label{b1eps}
\ee
then this means that:
\begin{enumerate}
\item One is assuming the phases $\alpha_+$ and $\alpha_-$ to be equal.
\item One is fixing
\be
\epsilon = \frac{\delta + i \sqrt{1 - \delta^2} \sin{\theta}}
{1 + \sqrt{1 - \delta^2} \cos{\theta}}.
\label{definition of epsilon}
\ee
\end{enumerate}
This is in fact what BB have {\em implicitly} done.

It follows from eqn~(\ref{definition of epsilon}) that
\be
\frac{2 \epsilon}{1 + \left| \epsilon \right|^2} =
\delta + i \sqrt{1 - \delta^2} \sin{\theta}.
\ee
Then,
both $2 \real\, \epsilon / \left( 1 + \left| \epsilon \right|^2 \right)
= \delta$
and $2 \imag\, \epsilon / \left( 1 + \left| \epsilon \right|^2 \right)
\approx \sin{\theta}$
are measurable CP-violating parameters.
In this sense,
it is true that ``both $\real\, \epsilon$ and $\imag\, \epsilon$
are observable quantities'',
as BB have written;
on the other hand,
$\epsilon$ cannot really be considered
a phase-convention-independent parameter,
because its definition depends on the phase convention $\alpha_+ = \alpha_-$.
Also,
$\epsilon$,
besides being phase-convention-dependent---it depends
on the relative phase of $|B_+\rangle$ and $|B_-\rangle$---,
is a completely artificial parameter,
for it joins together $\delta$,
which only depends on the mixing of $B_d$ and $\bar B_d$,
and $\theta$,
which depends on the specific initial state $B_+$
used in a particular experiment.

Therefore,
BB's assertion that $\imag\, \epsilon$ represents a new form of CP violation
{\em in the mixing} of $B_d$ and $\bar B_d$ is wrong.
BB have been taken to believe this because,
instead of working
with an arbitrary $B_+$---an arbitrary $\zeta$---to begin with,
they have assumed $B_+$ to be an eigenstate of CP.
Unfortunately,
as we shall see in section~\ref{seccao final},
that assumption cannot be realized in a concrete experiment.

\section{The phase of $q/p$}

Up to now,
the phase $\zeta$ has not been specified and,
as such,
$\theta$ is free too.
In their paper,
BB have specifically suggested using as initial states eigenstates of CP.
Thus,
if
\be
\begin{array}{rcl}
{\cal C} {\cal P} |B_d\rangle \!\!\!&=&\!\!\! e^{i \xi} |\bar B_d\rangle,
\\*[1mm]
{\cal C} {\cal P} |\bar B_d\rangle \!\!\!&=&\!\!\! e^{- i \xi} |B_d\rangle,
\end{array}
\label{CP dos kets}
\ee
then BB would want to use $\zeta = \xi$.
Clearly,
$B_+$ is then the CP-even eigenstate and $B_-$ is the CP-odd eigenstate.

For this very specific choice,
BB have proceeded to compute $\theta$.
In order to do this they have computed the phase of $q/p$.
Assuming $\left| \Gamma_{12} \right| \ll \left| M_{12} \right|$,
one has \cite{silva}
\be
\frac{q}{p} = \sqrt{\frac{M_{12}^\ast}{M_{12}}}.
\label{merdacp1}
\ee
The matrix element $M_{12}$ is given by the standard-model box diagram,
which is dominated by intermediate top quarks.
One obtains
\be
\frac{M_{12}^\ast}{M_{12}} = \frac
{\left( V_{tb}^\ast V_{td} \right)^2 \langle \bar B_d|
\left[ \bar b \gamma^\mu \left( 1 - \gamma_5 \right) d \right]
\left[ \bar b \gamma_\mu \left( 1 - \gamma_5 \right) d \right]
|B_d \rangle}
{\left( V_{tb} V_{td}^\ast \right)^2 \langle B_d|
\left[ \bar d \gamma_\mu \left( 1 - \gamma_5 \right) b \right]
\left[ \bar d \gamma^\mu \left( 1 - \gamma_5 \right) b \right]
|\bar B_d \rangle}.
\ee
The matrix elements may be related to each other by means of the CP symmetry
of the strong interactions.
In order to do this one must use,
besides eqns~(\ref{CP dos kets}),
the CP transformation of the quark fields,
which reads
\be
\begin{array}{rcl}
\left( {\cal C} {\cal P} \right) d \left( {\cal C} {\cal P} \right)^\dagger
\!\!\!&=&\!\!\!
e^{i \xi_d} \gamma^0 C \bar d^T,
\\*[1mm]
\left( {\cal C} {\cal P} \right)
\bar b \left( {\cal C} {\cal P} \right)^\dagger
\!\!\!&=&\!\!\!
- e^{- i \xi_b} b^T C^{-1} \gamma^0,
\end{array}
\ee
where $\xi_d$ and $\xi_b$ are {\em arbitrary CP-transformation phases}.
It follows that
\be
\frac{q}{p} = \pm \frac{V_{tb}^\ast V_{td}}{V_{tb} V_{td}^\ast}
e^{i \left( \xi + \xi_d - \xi_b \right)}.
\label{qoverp}
\ee
Therefore,
\ba
\theta \!\!\!&=&\!\!\! \xi - \arg{\frac{q}{p}}
\no    \!\!\!&=&\!\!\! 2 \arg{\left( V_{tb} V_{td}^\ast \right) }
+ \xi_b - \xi_d\ (\mbox{mod}\ \pi).
\label{true theta}
\ea
This is exactly the path followed by BB,
with one important exception:
{\em BB omitted the arbitrary phases $\xi_b$ and $\xi_d$}
in the CP transformation of the quark fields,
implicitly setting them to zero.
They obtained $\theta = 2 \arg{\left( V_{tb} V_{td}^\ast \right) } + \pi$,
which depends on the phases chosen for the $b$- and $d$-quark fields.
But,
as $\theta$ is an observable phase,
it must be rephasing-invariant.
Thus,
it is clear that BB's procedure is meaningless.

One is not allowed to light-heartedly discard the arbitrary phases
$\xi_b$ and $\xi_d$.
The CP-transformation phases of the quark fields are essential,
as may be seen for instance
when one sets out to study the CP-invariance conditions
for the charged-current Lagrangian
\be
\frac{g}{2 \sqrt{2}} \sum_{\alpha = u, c, t}
\sum_{k = d, s, b} \left[ W_\mu^+ V_{\alpha k} \bar \alpha \gamma^\mu
\left( 1 - \gamma_5 \right) k
+ W_\mu^- V_{\alpha k}^\ast \bar k \gamma^\mu
\left( 1 - \gamma_5 \right) \alpha \right].
\ee
The most general CP transformation is
\be
\begin{array}{rcl}
{\displaystyle \left( {\cal C} {\cal P} \right)
W_\mu^+ \left( {\cal C} {\cal P} \right)^\dagger}
\!\!\!&=&\!\!\!
{\displaystyle - e^{i \xi_W} W^{\mu -},}
\\*[1mm]
{\displaystyle \left( {\cal C} {\cal P} \right)
\bar \alpha \left( {\cal C} {\cal P} \right)^\dagger}
\!\!\!&=&\!\!\!
{\displaystyle - e^{- i \xi_\alpha} \alpha^T C^{-1} \gamma^0,}
\\*[1mm]
{\displaystyle \left( {\cal C} {\cal P} \right)
k \left( {\cal C} {\cal P} \right)^\dagger}
\!\!\!&=&\!\!\!
{\displaystyle e^{i \xi_k} \gamma^0 C \bar k^T.}
\end{array}
\ee
If there is to be CP invariance,
the CP-transformation phases must be chosen such that
\be
V_{\alpha k} = e^{i \left( - \xi_W + \xi_\alpha - \xi_k \right)}
V_{\alpha k}^\ast.
\label{afinal}
\ee
Clearly,
if it was not for the freedom
allowed by the phases $\xi_\alpha$ and $\xi_k$,
it would suffice that any two elements of $V$
have different phases for CP to be violated.
It is well known that things are not so:
a whole quartet $V_{\alpha k} V_{\beta j} V_{\alpha j}^\ast V_{\beta k}^\ast$
must be non-real in order for there to be CP violation.

\section{A practical experiment} \label{seccao final}

From the previous section,
and in particular from eqn~(\ref{true theta}),
one gathers that {\em the phase $\xi - \arg{q/p}$ is not measurable}.
On the other hand,
one knows from eqn~(\ref{a+cp(t)})
that {\em the phase $\zeta - \arg{q/p}$ is measurable}.
One can only conclude that
{\em $\zeta$ can never be equal to $\xi$ in a real experiment}.

The phase $\zeta$ in the initial state $B_+$ must be such that:
\begin{enumerate}
\item It includes the CP-transformation phase $\xi + \xi_d - \xi_b$,
which is going to cancel out a similar term in $\arg{q/p}$---see
eqn~(\ref{qoverp}).
\item It includes the phase of some elements of $V$,
in such a way that $\zeta - \arg{q/p}$
is invariant under a rephasing of the quark fields.
\end{enumerate}
Taking $\zeta = \xi$ does not satisfy the above conditions.
This means that the suggestion by BB,
that the initial state $B_+$ be taken to be a CP eigenstate,
is {\em unrealizable in practice}.

In order to convince oneself of this fact,
one may consider a specific set-up
for an experiment of the kind suggested by BB.
Suppose that one wanted the initial state to be the CP-odd eigenstate $B_-$.
At a $B$-factory one uses the decay of the resonance $\Upsilon (4S)$
to produce a $B_d \bar B_d$ pair in an antisymmetric state;
if at a certain instant the meson in the left side of the detector
is observed to decay into a CP-even state,
we may presume that the meson in the right side of the detector is,
at that instant,
$B_-$.

One must however be careful and study in detail
the decay into the specific CP-even state that one uses as a tag.
That CP-even state may be,
for instance,
$\pi^+\pi^-$.
Now,
the linear combination of $|B_d\rangle$ and $|\bar B_d\rangle$
which decays into $\pi^+ \pi^-$ is
\be
|B_{\rm yes}\rangle = \langle \pi^+\pi^-|T|B_d\rangle^\ast |B_d\rangle
+ \langle \pi^+\pi^-|T|\bar B_d\rangle^\ast |\bar B_d\rangle;
\ee
indeed,
the orthogonal linear combination,
\be
|B_{\rm no}\rangle = \langle \pi^+\pi^-|T|\bar B_d\rangle |B_d\rangle
- \langle \pi^+\pi^-|T|B_d\rangle |\bar B_d\rangle
\ee
clearly cannot decay into $\pi^+\pi^-$.
Thus,
if at a certain instant we observed $\pi^+\pi^-$
in the left side of the detector,
we would know the meson in the right side of the detector to be,
at that instant,
not $B_-$ as we might presume,
but rather $B_{\rm no}$.
This means that
\be
e^{i \zeta} = - \frac{\langle \pi^+\pi^-|T|B_d\rangle}
{\langle \pi^+\pi^-|T|\bar B_d\rangle}.
\ee
Assuming the decays to be given by the standard-model tree-level diagrams,
we would have
\ba
e^{i \zeta}
\!\!\!&=&\!\!\! - \frac
{V_{ub}^\ast V_{ud} \langle \pi^+\pi^-|
\left[ \bar b \gamma^\mu \left( 1-\gamma_5 \right) u \right]
\left[ \bar u \gamma_\mu \left( 1-\gamma_5 \right) d \right]
|B_d\rangle}
{V_{ub} V_{ud}^\ast \langle \pi^+\pi^-|
\left[ \bar u \gamma_\mu \left( 1-\gamma_5 \right) b \right]
\left[ \bar d \gamma^\mu \left( 1-\gamma_5 \right) u \right]
|\bar B_d\rangle}
\no
\!\!\!&=&\!\!\! - \frac{V_{ub}^\ast V_{ud}}{V_{ub} V_{ud}^\ast}
e^{i \left( \xi + \xi_d - \xi_b \right) },
\ea
because the CP-parity of $\pi^+\pi^-$ is $+1$.
We would thus obtain
\be
\theta = \zeta - \arg{\frac{q}{p}} =
2 \arg{\left( V_{ud} V_{tb} V_{ub}^\ast V_{td}^\ast \right) }\
(\mbox{mod}\ \pi).
\ee
As expected,
$\theta$ is independent of the CP-transformation phases
and is rephasing-invariant.

\section{Conclusions}

I conclude that BB's claimed discoveries are spurious.
Instead of talking loosely about using CP-eigenstate initial states,
it is essential to take into account
the exact physical mechanism that one uses to tag the initial state.
There is no new form of indirect CP violation,
contrary to what BB have claimed---$\imag\, \epsilon$ is a CP-violating
parameter in the relationship between mixing and the decay amplitudes;
the latter originate in the tagging of the initial state.
The construction itself of the parameter $\epsilon$ is artificial
and devoid of any physical meaning.

\end{document}